\documentclass{emulateapj}
\usepackage{graphicx}

\shorttitle{Nucleosynthesis in the outflows associated with accretion disks of Type II collapsars}
\shortauthors{Banerjee \& Mukhopadhyay}

\begin{document}

\title{Nucleosynthesis in the outflows associated with accretion disks of Type II collapsars}

\author{Indrani Banerjee\altaffilmark{1} and Banibrata Mukhopadhyay\altaffilmark{1}} 
\altaffiltext{1}{Department of Physics, Indian Institute of Science, Bangalore 560 012, India;\\
indrani@physics.iisc.ernet.in , bm@physics.iisc.ernet.in}

\begin{abstract}
We investigate nucleosynthesis inside the outflows from gamma-ray burst (GRB) accretion disks formed by the 
Type II collapsars. In these collapsars, massive stars undergo core collapse to form a proto-neutron star
initially and a mild supernova explosion is driven. 
The supernova ejecta 
lack momentum and subsequently this newly formed neutron star gets transformed to a stellar mass 
black hole via massive fallback. The hydrodynamics and the nucleosynthesis in these accretion disks has 
been studied extensively in the past. Several heavy elements are synthesized in the disk and much of these 
heavy elements are ejected from the disk via winds and outflows. We study nucleosynthesis in 
the outflows launched from these disks by using an adiabatic, spherically expanding 
outflow model, to understand which of these elements thus synthesized in the disk survive in the 
outflow. While
studying this we find that many new elements like isotopes of titanium, copper, zinc etc. 
are present in the outflows. 
$^{56}$Ni is abundantly synthesized in most of the cases in the outflow 
which implies that the outflows from these disks in a majority of cases will lead to an observable  
supernova explosion. It is mainly present when outflow is considered from the He-rich, $^{56}$Ni/$^{54}$Fe
rich zones of the disks. However, outflow from the Si-rich zone of the disk remains rich in silicon.
Although, emission lines of many of these heavy elements have been observed in the X-ray afterglows of 
several GRBs by Chandra, BeppoSAX, XMM-Newton etc., Swift seems to have not detected these lines yet.

\end{abstract}


\section{Introduction}
Over the last two decades the availability of data for 
studying long-duration gamma-ray bursts (GRBs) has drastically increased. Long-duration bursts last from 
two seconds to several 
minutes, while short duration bursts range from a few milliseconds to two seconds. It is believed that they 
are caused by completely different physical phenomena. Although the driving mechanisms behind the 
long-duration bursts are well-understood by now, the underlying theories governing the short-duration 
bursts are 
still in their infancy. Collapsar model (Woosley 1993; MacFadyen \& Woosley 1999, hereafter MW99, 
MacFadyen et al. 2001) is one of the most promising theoretical 
models 
explaining the core collapse of massive stars to accretion disks surrounding stellar mass black holes that 
drive jets of material to blast outward perpendicular to the disk causing long-duration GRBs.

GRBs were discovered accidentally in the late 1960s by the Vela satellites launched by 
the USA, to keep track of the fact that the nuclear test ban treaty is strictly followed by the U.S.S.R.
and other nations. However, the results from 
the Vela satellites were published several years later (Klebesadel et al. 1973), which were 
quickly confirmed by data from the Soviet Konus satellites (Mazets et al. 1974). Ever since 
the detection of the first GRBs, their nature and origin remained a subject of great interest to the
astronomers. Besides gamma rays, afterglows were observed in the X-ray, optical and radio wavelengths which  
enormously helped to establish a working hypothesis of the GRBs. Many observations in the 
recent years reveal that several GRBs occur in coincidence with core-collapse supernovae (SNe) 
(M\'{e}sz\'{a}ros 2002). A turning point in the 
history of GRB research arose with the discovery of GRB 980425 in conjunction with one of the most unusual 
SNe ever seen, SN1998bw (Galama et al. 1998). After that many more such events followed, namely, the 
association 
of SN 2003dh with GRB 030329 (Hjorth et al. 2003), SN 2003lw with GRB 031203 and SN 2002lt with
GRB 021211; (see, Woosley \& Bloom 2006; Piran 2005; Matheson 2005; H\"{o}flich et al. 2005; 
Della Valle 2005; Bloom 2005; Nomoto et al. 2004). Most of the above mentioned supernovae exhibited 
strong oxygen lines and possessed energies several orders of magnitude larger than typical supernovae
and hence they are termed as `hypernova'. Nevertheless, 
SN 2006aj in conjunction with GRB060218 appears to be
an exception as it exhibited a dimmer SN with weak oxygen lines and X-Ray Flashes (XRF) (Nomoto et al. 
2010).

However, all SNe of Type Ibc, which occur due to the core collapse of very massive stars after losing their 
hydrogen and/or helium envelope, are not associated with GRBs. 
Similarly, all long-duration GRBs are not associated with supernovae e.g. GRB 060614 (Zhang 2006).
GRBs result from most massive and most 
rapidly rotating low metallicity stars. This is because only very massive stars can undergo core collapse
to form black holes. The rapid rotation endows the star the necessary angular momentum so that an accretion 
disk develops around the black hole and low metallicity reduces mass loss 
(and hence angular momentum loss) from the progenitor (Heger et al. 2003; MacFadyen \& Woosley 1999;
Woosley \& Heger, 2006).

Collapsar models which explain the formation of accretion disks around stellar mass black holes as a 
result of core collapse of rapidly rotating massive stars can be categorized broadly into two kinds, 
namely, Type I and Type II collapsars. In Type I collapsar, the star undergoing core collapse is very 
massive with mass in the main sequence $(M_{MS})$ at least greater than $40 M_{\odot}$. When these stars
undergo core collapse they directly form stellar mass black holes. No observed supernova explosion is 
associated with such core collapse events and hence they are often termed as ``failed supernovae" 
(Fryer 1999). If 
the progenitor star has sufficient angular momentum (MW99), which is in a reasonable  
range of progenitor model predicted by Heger et al. (2000), an 
accretion disk develops around the black hole. The disk stays for $10-20$ s 
(MW99) and during this time very high accretion rate $(\dot{M})$ with $\dot{M} > 0.1 M_{\odot} 
s^{-1}$ is maintained. Due to such high $\dot{M}$ the temperature of the disk rises 
significantly above $10^{10}$K,
specially in the inner region of the disk close to the black hole, which enables the photodisintegration of 
all the heavy nuclei (which might have been formed in the outer disk) into free nucleons. 
While nucleosynthesis is not so important in order to produce
heavy elements in such accretion disks itself (Banerjee \& Mukhopadhyay 2013a),
%, formed by Type I collapsars, 
there may be outflows from the neutron-proton rich regions (Surman et al. 
2011; Surman et al. 2006) where due to appropriate lowering of temperature and density, the free nucleons 
may recombine to give rise to $^{56}$Ni and other heavy elements.
Since, nucleosynthesis in the context of outflows from Type I collapsars has 
been studied extensively in the past (e.g., Surman et al. 2006; Surman et al. 2011 etc.),
we do not repeat it in the present context.

Type II collapsars are formed from the core collapse of progenitors with mass  
$\\20 M_{\odot} \lesssim M_{MS} < 40 M_{\odot}$. In this case, first a proto-neutron star is formed after 
the core 
collapse and a mild supernova explosion is driven. However, a strong reverse shock causes the supernova 
ejecta to fall back onto the newly formed neutron star which then immediately collapses into a black hole. 
The increase in progenitor mass enhances the amount of supernova ejecta falling back onto the nascent 
neutron star (Woosley \& Weaver 1995; Fryer 1999; Fryer \& Heger 2000). $\dot{M}$ is also 
lower in this case, $10^{-4}\rm ~to ~10^{-2}$ $M_\odot s^{-1}$ (Fujimoto et al. 2003), and this accretion
may continue for hundreds to thousands of seconds (MacFadyen et al. 2001). 
Lower $\dot{M}$ decreases the 
density and temperature in these accretion disks compared to collapsar~I disks which makes them 
ideal
sites for the nucleosynthesis of heavy elements. 
%Since our goal in the present work is to understand, how 
%heavy elements, more precisely, iron group elements that contaminate the galaxies, are synthesized, we 
%concentrate mainly on the Type II collapsars.   
%We report the synthesis of several unusual nuclei like $^{31}$P, $^{39}$K, $^{43}$Sc, $^{35}$Cl, 
%and various uncommon
%isotopes of titanium, vanadium, chromium, manganese and copper in the disk and also confirm that isotopes 
%of iron, cobalt, 
%nickel, argon, calcium, sulphur and silicon get synthesized in the disk, as shown by previous authors.

Nucleosynthesis in the accretion disks of Type II collapsars has already been studied in detail  
(Banerjee \& Mukhopadhyay 2013a,b). 
Here we examine nucleosynthesis in the 
outflows launched from them considering an adiabatic, one-dimensional and spherically expanding outflow
model \cite{Fujimoto04}. 
As matter accretes onto the central object, some fractions of the accreting gas could be ejected from the
disk through outflows and jets. Jets can be driven by neutrino processes (MW99; Fryer \& M\'{e}sz\'{a}ros 2003) and/or
magnetohydrodynamic (MHD) processes (Mizuno et al. 2004; Proga et al. 2003) with relativistic velocities
which exhibit as GRBs. 
It has been recently proposed that a large fraction of GRBs should have a magnetically driven jet, since
neutrino-driven jets have too much baryon loading (Lei et al. 2013).
Nucleosynthesis in the jets has been studied extensively in the past 
(Fujimoto et al. 2008, Fujimoto et al. 2007; Beloborodov 2003; Inoue et al. 2003; Lemoine 2002; Pruet et 
al. 2002). 
%As the jet propagates, $^{56}$Ni is abundantly synthesized.
$^{56}$Ni is abundantly synthesized in the jets. Additionally, $^{56}$Ni is also synthesized in copious
amounts in the outflows from the collapsar accretion disks (MW99; Pruet et al. 2003, Milosavljevi\'{c} et 
al. 2012,) and also in explosive burning as a shock wave propagates out through the star
(Maeda \& Nomoto 2003; Fryer et al. 2006; Maeda \& Tominaga 2009).
Apart from accreting black holes, protomagnetars which can serve as another possible central engine
for long-duration GRBs are also capable of producing neutron-rich outflows (Metzger et al. 2008).
MacFadyen \& Woosley (1999) showed that as collapsar disks accrete matter on to the black hole,
they also eject substantial amount of matter through outflows. These outflows often result in stellar
explosions which may be observable as Type Ibc supernovae (Kumar et al. 2008). 
SN1998bw and SN1997ef are examples of such
supernovae (MacFadyen 2003). 
It has been suggested that these explosions may take place independent of any jet resulting in GRBs. 
It is worth mentioning that a stellar explosion does
 not necessarily lead to a supernova. For a stellar explosion to be observable as a supernova, there 
has to be a continuous source of energy 
  to power the observable light curve for long times, of the order of weeks to months. The time evolution of
the light curve of Type Ibc supernovae results from the decay of $^{56}$Ni to $^{56}$Co which subsequently
 decays to $^{56}$Fe. The half-lives of these decays are 6 days and 77 days respectively. Once these nuclei undergo beta decay, their daughter nuclei emit gamma rays which eventually come in
 thermal equilibrium with the matter ejected in the outflow. The energy from these decays is later re-emitted in the optical band and observed as the supernova
 light curve. Hence the temporal evolution of
  the supernova light curve is directly dependent on the amount of $^{56} \rm Ni$ produced in the outflow.
   If sufficient amount of $^{56}$Ni is not produced in the outflow, the stellar explosion would still take
place but it would not be observable and hence there will be {\it no supernova}.

In this work we have found that $^{56}$Ni is abundantly synthesized in most of the cases in the outflow 
which signifies that the outflows from these disks in a majority of cases will lead to an observable  
supernova explosion.  
Apart from $^{56}$Ni, some other isotopes of nickel ($^{57}$Ni,  $^{58}$Ni) and $^{44}$Ti,
$^{60}$Zn, $^{62}$Zn and $^{59}$Cu are synthesized abundantly in the outflow. 
$^{28}$Si, $^{32}$S, $^{36}$Ar and $^{40}$Ca are also synthesized in the outflow in some of the cases.
By investigating nucleosynthesis in the outflow, we can conclusively say that which 
of the elements leave their signatures in the observation.
This is important because, emission lines of many of these heavy elements have 
been observed in the X-ray afterglows of several GRBs by BeppoSAX, Chandra, XMM-Newton etc., e.g. 
iron lines in GRB 970828 
(Yoshida et al. 1999), GRB 
970508 (Piro et al. 1999), and GRB 000214 (Antonelli et al. 2000); magnesium, silicon, 
sulphur, argon and 
calcium lines in GRB 011211 (Reeves at al. 2003); sulphur lines in GRB 991216 (Piro et al. 2000)
and GRB 020813 (Butler et al. 2003). 
%However, we should take these results with caution since none of these emission lines have been 
%detected by Swift yet (Zhang et al. 2006, Hurkett et al. 2008).
We also present the nuclear reactions responsible for the synthesis of 
such heavy elements.   

The paper is organized as follows. In the next section, we briefly mention the hydrodynamic model adopted 
to study the collapsar accretion disk and discuss the main results we have obtained while studying the 
nucleosynthesis in the accretion disk.
In \S 3 we give the details of the outflow model chosen for the
present purpose.
\S 4 is devoted to the nucleosynthesis in the different outflow models with 
the details of the underlying 
nuclear reactions and finally we end with a summary and discussion in \S 5.

\section{Disk model and input physics}

We adopt height-averaged equations based on a pseudo-Newtonian framework \cite{Mukhopadhyay}.
The accretion disk formed in a Type II collapsar is modelled within the 
framework suggested by previous authors \cite{Kohri, Chen} where the electron degeneracy pressure 
and the evolving neutron to proton ratio are appropriately calculated. 
The details of the hydrodynamics and the nucleosynthesis in the disk are described by Banerjee \& 
Mukhopadhyay (2013a,b). Here we recall some of the salient features of the nucleosynthesis in the disk
for completeness.
%The equations that we have used
%and the assumptions that we have taken are described in detail by Banerjee \& Mukhopadhyay (2013).
Table 1 describes the various disk models we use for our work. 
While analysing the nucleosynthesis in the disk we choose that the
disk is once He-rich and then it is Si-rich in the outer region. 
In Banerjee \& Mukhopadhyay (2013a,b), we have reported synthesis of several unusual nuclei like $^{31}$P, 
$^{39}$K, $^{43}$Sc, $^{35}$Cl and various uncommon
isotopes of titanium, vanadium, chromium, manganese and copper in the disk, apart from 
isotopes of iron, cobalt, nickel, silicon, sulphur, argon and calcium which have already been reported 
earlier.   
Moreover, analysing the disk results we have found that several zones, characterized by dominant 
elements, are formed in the disk. For example in the He-rich disk the region between 
$1000-300R_g$ is rich in $^{40}$Ca, $^{44}$Ti and $^{48}$Cr. Inside this region between $300-80R_g$ there 
is a zone which is overabundant in $^{56}$Ni, $^{54}$Fe, $^{28}$Si and $^{32}$S. Finally inside $80R_g$
all the heavy elements get photodisintegrated to $^{4}$He and some free nucleons are also synthesized. For 
the Si-rich disk the outermost zone from $1000-250R_g$ is rich in $^{28}$Si and $^{32}$S. Inside this 
region there is a narrow zone rich in $^{54}$Fe and $^{56}$Ni which extends 
upto $70R_g$ and finally the innermost region of the disk is again overabundant in $^4$He with the 
presence of some free nucleons. 
It is to be noted here that by overabundant we mean the
mass fractions of the elements like $^{56}$Ni, $^{54}$Fe, $^{28}$Si, $^{32}$S and $^{4}$He are
many orders of magnitude higher compared to their initial mass fractions 
at the outer disk. The elemental distribution 
in the He- and Si-rich disks
starts appearing identical once threshold density and temperature are reached irrespective of the initial
abundances. On increasing $\dot{M}$ the respective zones shift outwards. 
Underlying nuclear reactions were also described in detail by Banerjee \& Mukhopadhyay (2013a). 
\begin{table}[h]
\vskip0.2cm
{\centerline{\large Table 1}}
{\centerline{Various disk models}}
{\centerline{}}
\begin{center}
\begin{tabular}{|c|c|c|}

\hline
$\rm Model$ & $\dot{M}$ &  $\rm Initial ~ abundance$\\
\hline

%\multirow{3}{*}{2} & 1 & 1.5 & 0.066 \\
 $\rm I$ & $0.001$ & $\rm He-rich$\\
\hline
$\rm II$ & $0.001$  & $\rm Si-rich$ \\

\hline
$\rm III$ & $0.01$ & $\rm He-rich$ \\ 

\hline
$\rm IV$ & $0.01$ & $\rm Si-rich$ \\ \hline

\end{tabular}

\end{center}
\end{table}

%\begin{table}[ht]
%\centering % used for centering table
%\caption{Various disk models} % title of Table
%\begin{tabular}{c c c c} % centered columns (4 columns)
%\hline\hline %inserts double horizontal lines
%Model & $\dot{M}$ & Initial abundance \\ [0.5ex] % inserts table
%heading
%\hline % inserts single horizontal line
%I & 0.001 & He-rich  \\ % inserting body of the table
%II & 0.001 & Si-rich \\
%III & 0.01 & He-rich  \\
%IV & 0.01 & Si-rich  \\

%\hline %inserts single line
%\end{tabular}
%\label{table:one} % is used to refer this table in the text
%\end{table}

\section{Outflows from accretion disks of Type II collapsars}

The production of outflows depends on the degree to which the disk is cooled by neutrino emission and
photodisintegration of heavy nuclei. Since $\dot{M}$ is very high, it is always possible that
the matter may get deposited onto the accretion disk which favors outflow unless cooling plays an
important role to aid accretion. If cooling is efficient, then most of the incoming gas will be accreted,
else it will be ejected from the system in the form of outflows. The cooling processes depend on 
temperature
which in turn depend on the angular momentum of the stellar progenitor 
(MacFadyen 2003). The virial temperature and the angular momentum at the Keplerian radius are
inversely proportional to each other. The production of a supernova corresponding to a GRB thus depends on
the angular momentum of the disk. If the angular momentum is low enough so that the virial temperature
reaches above $10^{10}$K, then all the heavy elements will be photodisintegrated to free nucleons which
will initiate the pair capture reactions onto the free protons and neutrons which in turn will lead to the
emission of neutrinos. Thus, subsequently the system will be cooled efficiently. Hence, low angular 
momentum progenitors
may produce a GRB but not an observable supernova. When the angular momentum is large enough so that the
virial temperature falls below $5 \times 10^{9} \rm K$, photodisintegration and pair capture reactions get
suppressed, as a result of which the matter remains heated up subject to be ejected from the disk in the
form of outflows. In the intermediate range, there is a partial 
photodisintegration; hence although
photodisintegration cools the system, cooling due to pair capture is still suppressed. Thus
photodisintegration aids accretion in this regime. 

The winds/outflows can be driven by various processes, e.g. neutrino process, mechanisms
related to viscosity and magnetic centrifugal
force. We first briefly describe the mechanism driving winds via each of the processes. When the 
proto-neutron stars are formed after the core collapse of their progenitors, a region of low density and 
high
entropy is formed behind the supernova ejecta which is steadily heated up by the neutrinos emanating from
the proto-neutron star. Due to the deposition of the heat, the ejecta gains kinetic energy which 
accelerates
the material in it resulting in an outflow known as the neutrino-driven wind (Fischer et al. 2010). However
the geometry of this process is quite different from the collapsar scenario (MW99). When the accretion
rate is  high mainly in the context of Type I collapsars, the inner regions of the accretion disk become
very hot to emit copious amounts of neutrinos. But as the neutrino luminosity increases, neutrino
annihilation also becomes efficient and large energy deposition can occur in the polar regions which 
creates a
pressure gradient that has a component away from the disk. This pressure gradient pushes the gas of
radiation and pairs which drives the outflow. Thus, close to the black hole neutrino annihilation may lead
to energy deposition which can drive polar relativistic outflows containing expanding bubbles of
radiation, pairs and baryons (MW99).

High differential rotation in the accretion disks leads to enhanced viscous interaction which heats up the
disk, raises its entropy and drives a wind off the disk. This explains how outflows can be driven
by viscosity. However,
since the temperature in these accretion disks is very high, the gas is completely ionized which
generates electric currents, that in turn produces magnetic fields. Galeev et al. (1979) showed that
rotational shear and convection which are present in these disks can amplify the seed magnetic field to
very large values, $B \gtrsim 10^{15}\rm G$. The magnetic field can subsequently drive an outflow 
transferring a part of the Poynting flux to matter (Blandford \&
Payne 1982). For further details of how winds can be driven by viscosity and MHD processes,
see, MW99 and Daigne \& Mochkovitch (2002) respectively. Also see Levinson (2006) where 
general relativistic MHD driven outflows were comprehensively studied in the context of GRBs.
Barzilay \& Levinson (2008) investigated nucleosynthesis in the above mentioned outflows.
         
\subsection{Formulation of the outflow model}
Abundance of various elements synthesized in the disk usually evolves in the outflow. 
%Various charged particles therein decay and respective capture processes lead to abundance change in the 
The change in the abundance not only depends on the initial conditions of the ejecta (e.g. density, 
temperature and abundance at the radius from where the outflow is launched), but also on the detailed 
hydrodynamics of the wind. 

Various outflow models have been suggested by several authors. For the present purpose, we choose a 
hydrodynamic model which is assumed to be adiabatic and freely expanding (Fujimoto et al. 2004),
which was already implemented in past for the purpose of nucleosynthesis in collapsars. 
 The temperature of the ejecta, based on this model, is given by
 \begin{equation}
T_{ej} (t)=T_0 \left(\frac{R_{ej}}{R_{ej}+v_{ej} t}\right)^{3(\gamma - 1)}.
\end{equation}
Here, $R_{ej}$ is the radius of the accretion disk at which the outflow is launched, $v_{ej}$ is the
 velocity of the ejecta at that radius which is assumed to be constant during the ejection, $T_0$ is the
temperature at $R_{ej}$ and $\gamma$ is the adiabatic index. $T_0$  at $R_{ej}$ is known from the disk
hydrodynamics. We calculate $v_{ej}$ from the ratio of the rate at which matter is ejected
from the disk to the matter accreted in. Let us assume that $\dot{M}_{ej}$ be the rate at which matter is
ejected from the disk. Then if we assume that the gas is spherically expanding,
 \begin{equation}
\dot{M}_{ej} =4 \pi r^2 \rho v_{ej},
\end{equation}  
where, 
\begin{equation}
r^2 =R_{ej}^2 + H_{ej}^2,
\end{equation}
$H_{ej}$ being the disk height at which the outflow is launched and $\rho$ the outflowing matter density.
On the other hand at $R_{ej}$, the rate at which the matter accretes is,
 \begin{equation}
\dot{M}_{acc} =4 \pi R_{ej} \rho H_{ej} v_R,
\end{equation}
when total accretion rate,
 \begin{equation}
\dot{M} =\dot{M}_{acc} + \dot{M_{ej}} = constant.
\end{equation} 
Thus,
\begin{equation}
\frac{\dot{M}_{ej}}{\dot{M}_{acc}} = \frac{v_{ej}}{v_R}\left(\frac{R_{ej}}{H_{ej}} + \frac{H_{ej}}
{R_{ej}}\right),
\end{equation} 
where $v_R$ is the radial velocity of the inflow at $R_{ej}$
We assume a given $\dot{M}_{ej} / \dot{M}_{acc}$ at the ejection radius of the accretion disk. The
radial velocity and the disk scale height are known from the disk hydrodynamics and hence $v_{ej}$ at
$R_{ej}$ can be calculated. 
Once $v_{ej}$ is calculated, the temperature profile can be obtained from
equation (1), where we choose $\gamma = 5/3$.

Adiabaticity ensures no entropy change during the propagation of the outflow.
The entropy per baryon, S, of the ejecta is given by (Pruet et al. 2003,
Qian \& Woosley, 1996)
\begin{equation}
\frac{S}{k} \approx 0.052 \frac{T^3_{\rm MeV}}{\rho_{10}} + 7.4 + ln \frac{T^{3/2}_{\rm MeV}}{\rho_{10}}
 \approx S_0 ,
\end{equation} 
where $S_0$ is the entropy per baryon of the accretion disk in units of $k$ at $R_{ej}$, $T_{\rm MeV}$ is 
the temperature
of the outflow in units of $\rm MeV$ and $\rho_{\rm 10}$ is the density of the outflow in units of 
$ 10^{10} \; \rm { g \; cm^{-3}} $. The first term on the right hand side of equation (7) is the
contribution to the entropy per baryon due to the relativistic particles like photons and 
electron-positron 
pairs, while the last two terms represent the contribution due to the heavy
  nonrelativistic particles. Assuming that outflows are unlikely from a very large distance of the 
black hole, we consider outflow from $R_{ej} \lesssim 200 R_g$ \cite{Fujimoto04} for our spherical outflow. 
The temperature profile is already obtained from equations (1) and (6). Therefore, assuming a fixed 
entropy 
for the outflow, we obtain the density profile from equation (7).
Once the hydrodynamics of the outflow is determined, the abundance evolution of various elements in the 
ejecta
can be obtained in the same way as it has been done for the disk (Banerjee \& Mukhopadhyay 2013a) such that 
the initial composition of the
ejecta to begin with is the same as that in the disk at $R_{ej}$.

In the following sections, the abundance evolution of various elements and the subsequent reactions leading
to change of abundances will be systematically discussed considering the possibility that the outflow may be launched 
from different radii of the disk with different velocities of ejection, and the entropy of the outflow 
being once same as that of the disk and once double the entropy of the disk.

\section{Investigation of abundance evolution in the outflow}

In order to study the abundance evolution in the outflow, we use well tested nuclear network code, which 
has been implemented for more than two decades by, e.g., Chakrabarti et al. (1987) and Mukhopadhyay \& Chakrabarti 
(2000) in the context of accretion disks. Cooper et al. (2006) used it to study superbursts on the neutron 
star surface. We have modified this code further by increasing the nuclear network and including reaction 
rates from the JINA Reaclib Database, https://groups.nscl.msu.edu/jina/reaclib/db/ \cite{Cybert}.
  
As already shown by Banerjee \& Mukhopadhyay (2013a), the disk has several zones characterized 
by dominant elements. We investigate 
abundance evolution in the outflow from all the zones in the 
accretion disks mentioned in Table I that lie within $R_{ej} \lesssim 200 R_{g}$. 
We focus our attention mainly on the outflows from disks with $\dot{M}=0.001M_{\odot}s^{-1}$
where initially the accreting matter was chosen to be once
He-rich and then Si-rich. Then we briefly mention the main results for the outflows 
from relatively high $\dot{M}$ disks, because outflows from those
disks do not yield any new result.
While considering the outflow from the disks, we must keep in mind that
the material may be heated as it leaves the disk due to viscous heating and/or neutrino energy deposition.
However, heating due to neutrino energy deposition is mainly important in the context of Type~I 
collapsar accretion disks.
Most of the heating occurs when the outflow is being launched from the disk, while it evolves almost
adiabatically during its way away from the disk \cite{Pruet04}.
However, in an MHD driven wind the entropy in the outflow remains almost similar to that
of the disk at the point of ejection as there is no efficient heating source inside the wind 
\cite{Fujimoto04}.
Therefore, we consider outflows for both the cases, 
when once the entropy in the outflow remains close to  
that of the disk and then the entropy in the outflow becomes double than that of the disk. 
In each of the high entropy and the low entropy cases we consider two $v_{ej}$, one 
$\sim 10^5 \rm cm~s^{-1}$ and the other $\sim 10^8 \rm cm~s^{-1}$, depending on the outflow rate 
with respect to the inflow rate \cite{Fujimoto04}. 
The major nucleosynthesis products obtained and the underlying nuclear 
reactions taking place in the outflows are discussed in the subsequent subsections.

\subsection{Outflow from the He-rich disk}

Here we consider the disk flow with pre-SN He-rich composition at the outer region, which 
surrounds a $3 M_\odot$ Schwarzschild
black hole accreting at $\dot{M}=0.001M_\odot s^{-1}$.

\subsubsection{Outflow from $40R_g$}
\begin{figure*}
%\vskip-5cm
   \centering
\includegraphics[scale=0.8]{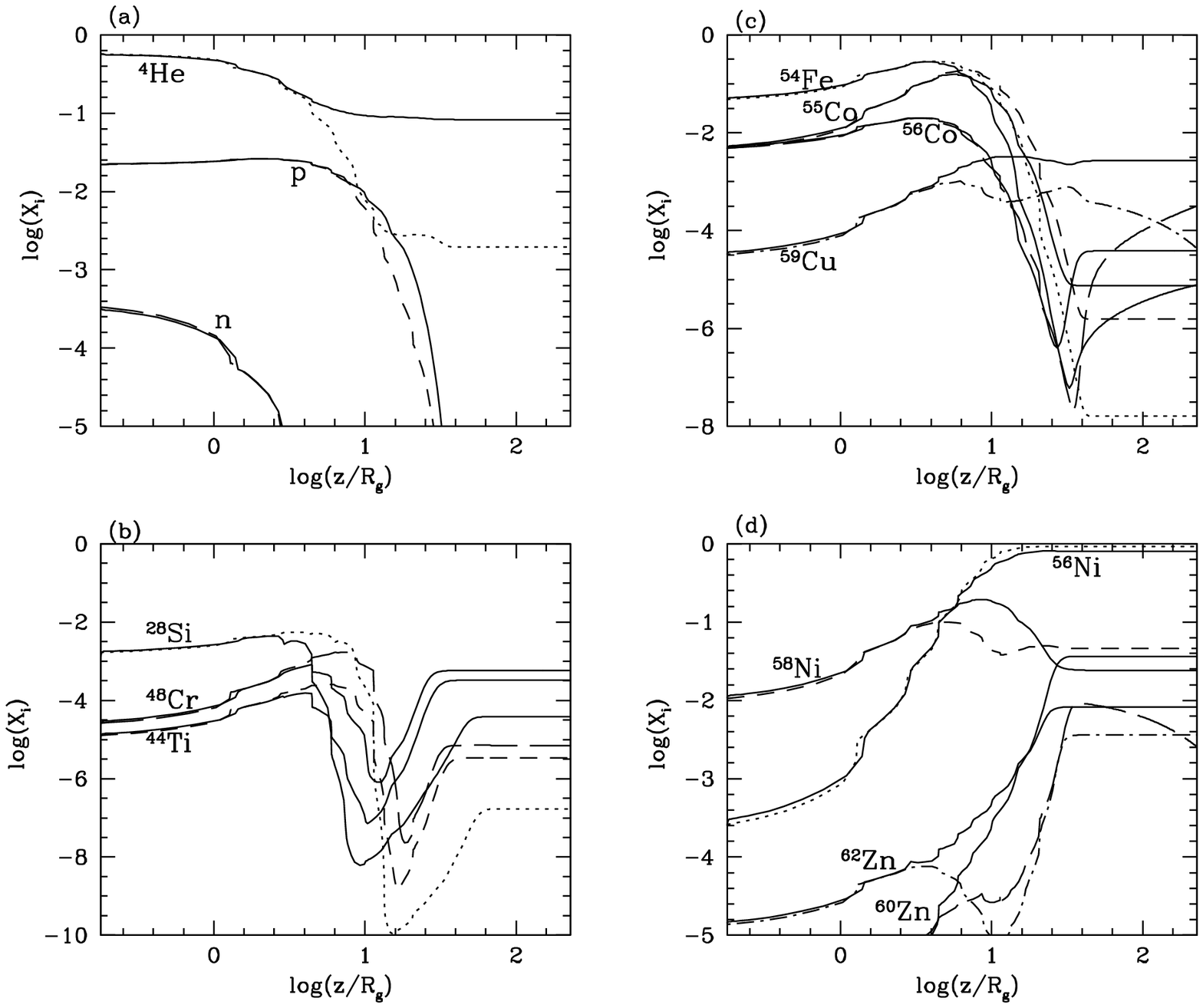}
\vskip-1.2cm
\caption{Abundance evolution in the ejecta while moving away from the disk, when outflow is being launched 
from $R_{ej} \sim 40 R_g $ of the accretion disk for a $3 M_\odot$ Schwarzschild black hole 
with $\dot{M}=0.001M_\odot s^{-1}$ with pre-SN He-rich abundance at the outer disk and the entropy of the 
ejecta being same as that of the disk. The variations of mass fraction of the 
elements shown by the solid lines correspond to $\dot{M}_{ej} / \dot{M}_{acc} = 100$ and the same elements 
when shown by linestyles other than solid correspond to
$\dot{M}_{ej} / \dot{M}_{acc} = 0.1$. }

\label{he40s}
\end{figure*}

\begin{figure*}
%\vskip-5cm
   \centering
\includegraphics[scale=0.8]{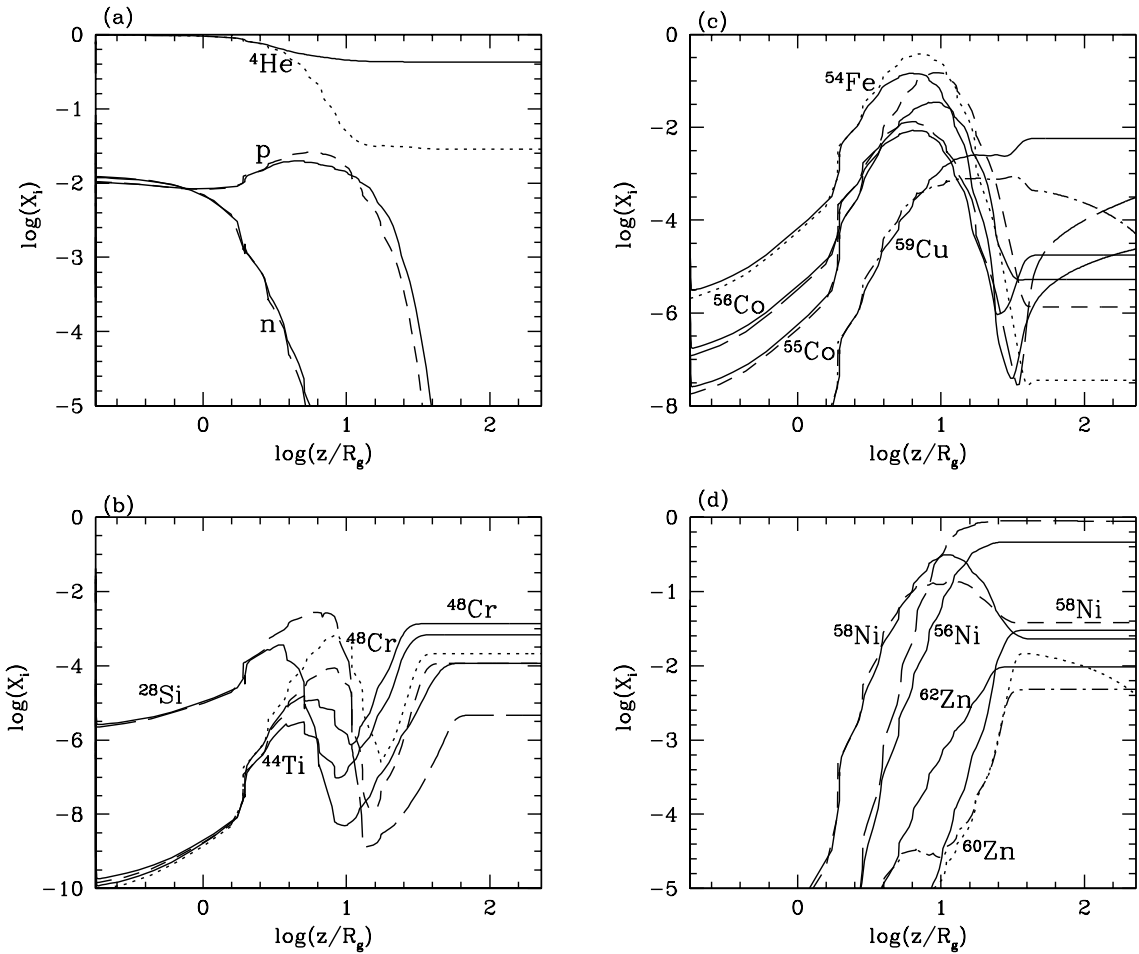}
\vskip-0.8cm
\caption{Same as in Fig. \ref{he40s}, except that the entropy of the ejecta being double than that 
of the disk.
} 
\label{he40s2}
\end{figure*}

\begin{figure*}
%\hskip-3cm
\vskip-3cm
\centering
\includegraphics[scale=0.8]{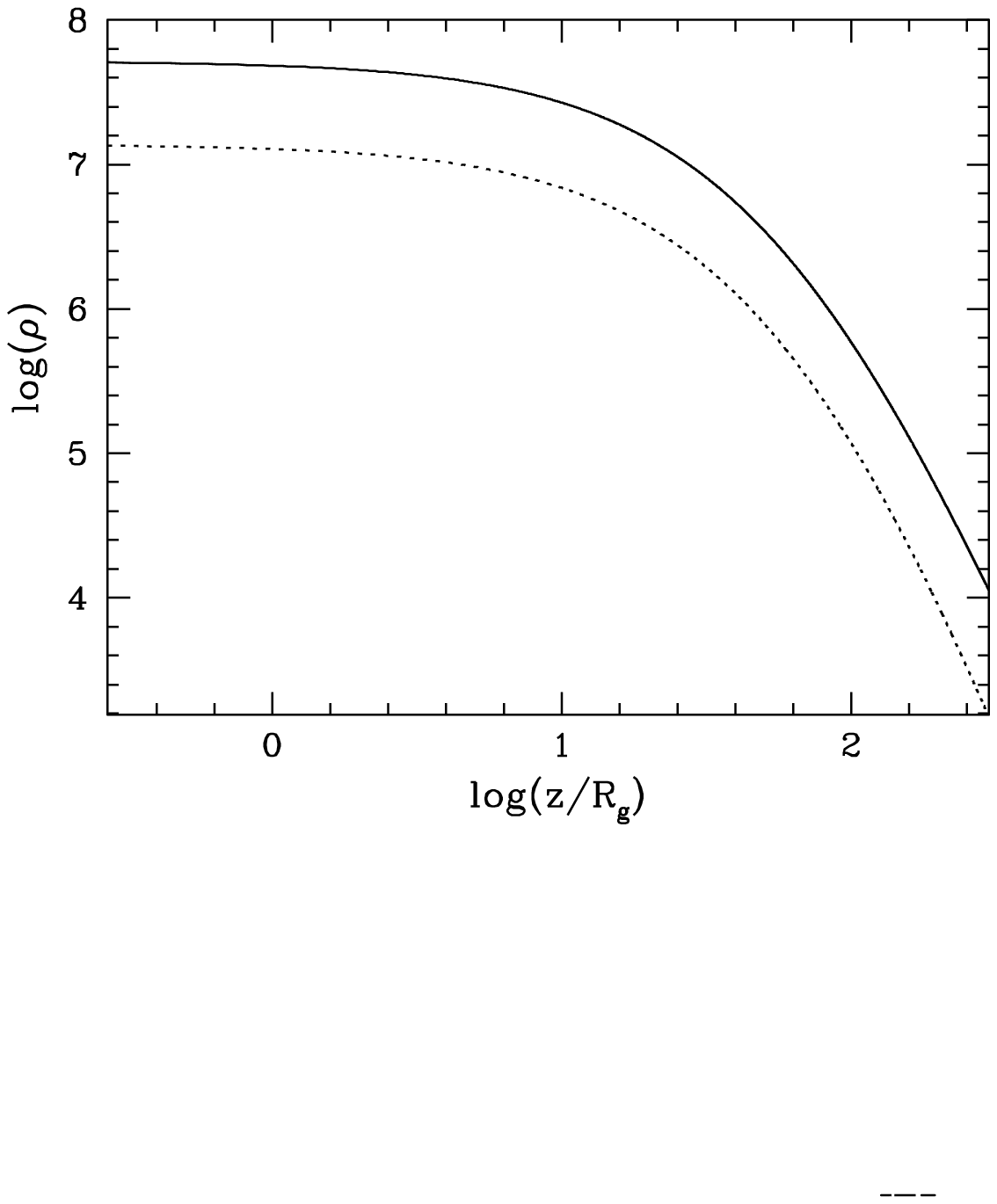}
\vskip-2.5cm
\caption{Typical density profiles of the outflow illustrating that the density increases on decreasing
the entropy of the outflow. The solid line corresponds to the case where the entropy is same as that of the
disk, while the dotted line denotes the case where the entropy is double than that of the disk.}
\label{dentem}
\end{figure*}

$R_{ej} \sim 40R_g$ lies primarily in the He-rich zone of the disk, but apart from $^4$He it also contains 
substantial amount of $^{54}$Fe. 
The entropy of the disk at $R_{ej} \sim 40R_g$ is $S \sim 15k$, which can be calculated from equation (7) 
knowing the underlying
disk hydrodynamics. The entropy of the outflow is fixed once to $S \sim 15k$ and then to $S \sim 30k$, 
and for each of these entropies we consider $\dot{M}_{ej} / \dot{M}_{acc} = 0.1$ and $100$ which correspond 
to $v_{ej} \sim 10^5 \rm cm~s^{-1}$ and $10^8 \rm cm~s^{-1}$ respectively.
With this information, the hydrodynamics of the outflow can be solved with which the abundance evolution 
can be studied, provided the initial composition of the ejecta is known. The initial composition 
is same as that in the accretion disk at $R_{ej}$. 

Figure 1 depicts the abundance evolution of the most important elements in the outflow from $R_{ej} \sim 
40R_g$
with the entropy of the outflow $S \sim 15k$. 
%The variations of mass fraction of various elements
%with distance from the accretion disk  when $\dot{M}_{ej} / \dot{M} = 100$ (i.e., $v_{ej}$ is high) is 
%shown by solid lines while the abundance evolution of elements with $\dot{M}_{ej} / \dot{M} = 0.1$ 
%(i.e., low $v_{ej}$) is depicted by linestyles other than solid. 
If we examine the abundances of $^4$He and $^{56}$Ni for both the high velocity and the low velocity cases 
in Fig. 1(a) and Fig.1(d) respectively, we notice that $^4$He has a lesser mass fraction but the 
yield of $^{56}$Ni in the outflow is higher
when
$v_{ej}$ is low.
This is because when $v_{ej}$ is low, it is easier for the $\alpha$-particles to recombine to form 
$^{56}$Ni.
Also, elements like $^{28}$Si, $^{32}$S, $^{36}$Ar, $^{40}$Ca, $^{44}$Ti and $^{48}$Cr experience a 
dip and then rise in their mass fraction while evolving in the outflow which is shown for $^{28}$Si, 
$^{44}$Ti and
$^{48}$Cr in Fig. 1(b). This characteristic behavior is observed in many more outflow cases and we are
reporting this for the first time in the literature.
In order to explain why this feature is observed in the mass fractions of the 
aforementioned elements, we have to understand the underlying reactions taking place in the outflow.
$^{28}$Si, $^{32}$S, $^{36}$Ar, $^{40}$Ca, $^{44}$Ti and $^{48}$Cr all react with excess $^4$He present in
the outflow to give rise to $^{56}$Ni. 
Apart from $^{56}$Ni, these $\alpha$-elements also give rise to some amount of $^{54}$Fe, $^{55}$Co, 
$^{56}$Co and $^{58}$Ni. 
The reactions are mainly a series of ($\alpha$,p) and ($\alpha$,n) 
reactions starting with the $\alpha$-elements. For example, $^{28}$Si gives rise to $^{56}$Ni via 
$^{28} \rm Si(\alpha,p)^{31}P(\alpha,p)^{34}S(\alpha,n)^{37}Ar(\alpha,p)^{40}K(\alpha,n)^{43}Sc(\alpha,p)\\
^{46}Ti(\alpha,n)^{49}Cr(\alpha,p)^{52}Mn(n,\gamma)^{53}Mn(p,n)^{53}Fe(\alpha,n)^{56}Ni$.
By means of similar reactions, $^{56}$Ni is obtained by starting with the other $\alpha$-elements as well.
%The nuclear reactions leading to the synthesis of the other elements will be elaborately discussed in a 
%future work which will emphasize more on the underlying nuclear physics.

$^{56}$Ni, which is the chief nucleosynthesis product in these cases, is obtained from two processes, 
first, 
from the direct recombination of $\alpha$-s and second from the ($\alpha$,p), ($\alpha$,n) 
reactions starting with the $\alpha$-elements. The density and velocity of the outflow decide
which out of the two processes plays the dominant role in the synthesis of $^{56}$Ni.
While both the above processes lead to a decrease in the abundance of $^4$He, the only process by which 
$^4$He can be synthesized is the recombination of neutrons and protons to $\alpha$-particles, which is also
not very effective in this case as the mass fraction of neutrons is quite different from that of 
protons \cite{Surman11}. Thus, the abundance of $\alpha$-particles declines quite rapidly in the outflow
(even more rapidly when the velocity of ejection is low).  Once the mass fraction of the $\alpha$-elements 
becomes lower than a certain threshold, the rates of ($\alpha$,p) and ($\alpha$,n) reactions decrease
significantly and their reverse reactions gain prominence to give rise to the $\alpha$-elements from
$^{54}$Fe, $^{55}$Co, $^{56}$Co and $^{58}$Ni. This marks the beginning of the rising arm of the $\alpha$-
elements.
For instance, $^{58}$Ni gives rise to $^{28}$Si
via, $^{58}\rm Ni (n,\alpha)^{55}Fe(p,\alpha)^{52}Mn(p,\alpha)^{49}Cr(n,\alpha)^{46}Ti(n,\alpha)\\^{43}Ca
(p,\alpha)^{40}K(p,\alpha)^{37}Ar(p,\alpha)^{34}Cl(p,\alpha)^{31}S(n,\alpha)^{28}Si$. 
Once the ($\alpha$,p), ($\alpha$,n) reactions and their inverses attain equilibrium, the mass fractions of 
the $\alpha$-elements become constant and then recombination of $^4$He is the only source of synthesizing
$^{56}$Ni. Production of $^{56}$Ni in copious amounts in the outflow is important because the temporal 
evolution of the light curve of core collapse supernovae is driven by the decay of $^{56}$Ni to $^{56}$Co
and subsequently to  $^{56}$Fe. Hence, synthesis of $^{56}$Ni in the outflow signifies that there will be 
an observable supernova explosion.

Apart from $^{56}$Ni, other isotopes of nickel ($^{57}$Ni, $^{58}$Ni), and $^{59}$Cu, $^{60}$Zn and 
$^{62}$Zn are also synthesized 
in copious amounts in the outflow which is illustrated in Fig. 1. 
Similar results were also obtained by Pruet et al. (2004b). 
Surman et al. (2006) studied nucleosynthesis in the outflow from GRB accretion 
disks and obtained
several unusual nuclei.  Kizivat et al. (2010) also obtained several light p-nuclei in the outflows from
GRB accretion disks. However, all of them investigated nucleosynthesis in the outflows associated with
Type~I collapsar accretion disks.
When $v_{ej}$ is low, the yields of
$^{59}$Cu, $^{60}$Zn and $^{62}$Zn are also lower compared to the high $v_{ej}$ case (see Figs. 1(c) and
1(d)). In fact, as the outflow advances, $^{59}$Cu and $^{60}$Zn are found to have a decreasing trend in 
the 
mass fraction when $v_{ej}$ is low.
This is because the reactions like $^{59} \rm Cu(n,\alpha) ^{56}Co$ and
$^{60}\rm Zn (\gamma,\alpha)^{56}Ni(n,\alpha)^{53}Fe(\alpha,p)^{56}Co$ are favored in the low $v_{ej}$ 
case. Thus, we also notice gradual rise in the mass fraction of $^{56}$Co as the ejecta evolves away
from the accretion disk which is evident from the long-dashed line of Fig. 1(c). 
%One of the possible ways of obtaining $^{58}$Ni in the outflow is from $^{28}$Si via, 
%$^{28} \rm Si(\alpha,p)^{31}P(\alpha,p)^{34}S(\alpha,n)^{37}Ar(\alpha,p)^{40}K(\alpha,n)^{43}Sc(\alpha,p)
%^{46}Ti(\alpha,n)^{49}Cr(\alpha,p)^{52}Mn(\alpha,p)^{55}Fe(\alpha,\gamma)\\^{59}Ni(\gamma,n)^{58}Ni$.

Figure 2 illustrates the abundance evolutions of the important elements in the ejecta when the outflow has 
double the entropy of the disk. 
On increasing the entropy of the outflow, the density of the ejecta declines (which is illustrated in 
Fig. 3) and hence there are lesser 
chances for the direct recombination of the $\alpha$-s to $^{56}$Ni.
%Here also the solid lines represent the abundance evolution of elements
%when $v_{ej}$ is high and the linestyles other than solid represent the evolution of mass fraction of
%elements for the low $v_{ej}$ case.
However, we assume, in the framework of the outflow model under consideration, that with the decrease
of density at the launching radius of the disk, the base of the outflow 
expands keeping it isothermal (maintaining $T_0$ to be the same as that with lower entropy). 
In such a case, the abundances of most of the heavy nuclei experience a sharp decline compared to the low entropy 
case in Fig. 1 when the outflow is being launched. Since the mass fractions of neutrons and protons are 
very similar in the outflow to begin
with (see Fig. 2(a)), they efficiently combine to give rise to $\alpha$-particles \cite{Surman11} which 
explains the 
rise in the mass fraction of $^4$He during the launching of the outflow. Note that $^4$He was
also not destroyed in the beginning of launching in the high entropy case. However, the mass fractions of 
all the heavy elements start 
rising thereafter and the trough-like feature, which was there for the $\alpha$-elements in the previous
figure, is also seen here. In fact, Fig. 1 and Fig. 2 have qualitatively the same features which implies 
that
the underlying nuclear physics taking place in both the cases is similar. Thus, to avoid repetition,
we are not going into the detailed description of Fig. 2. 

\subsubsection{Outflow from $120R_g$}
\begin{figure*}
%\vskip-5cm
   \centering
\includegraphics[scale=0.8]{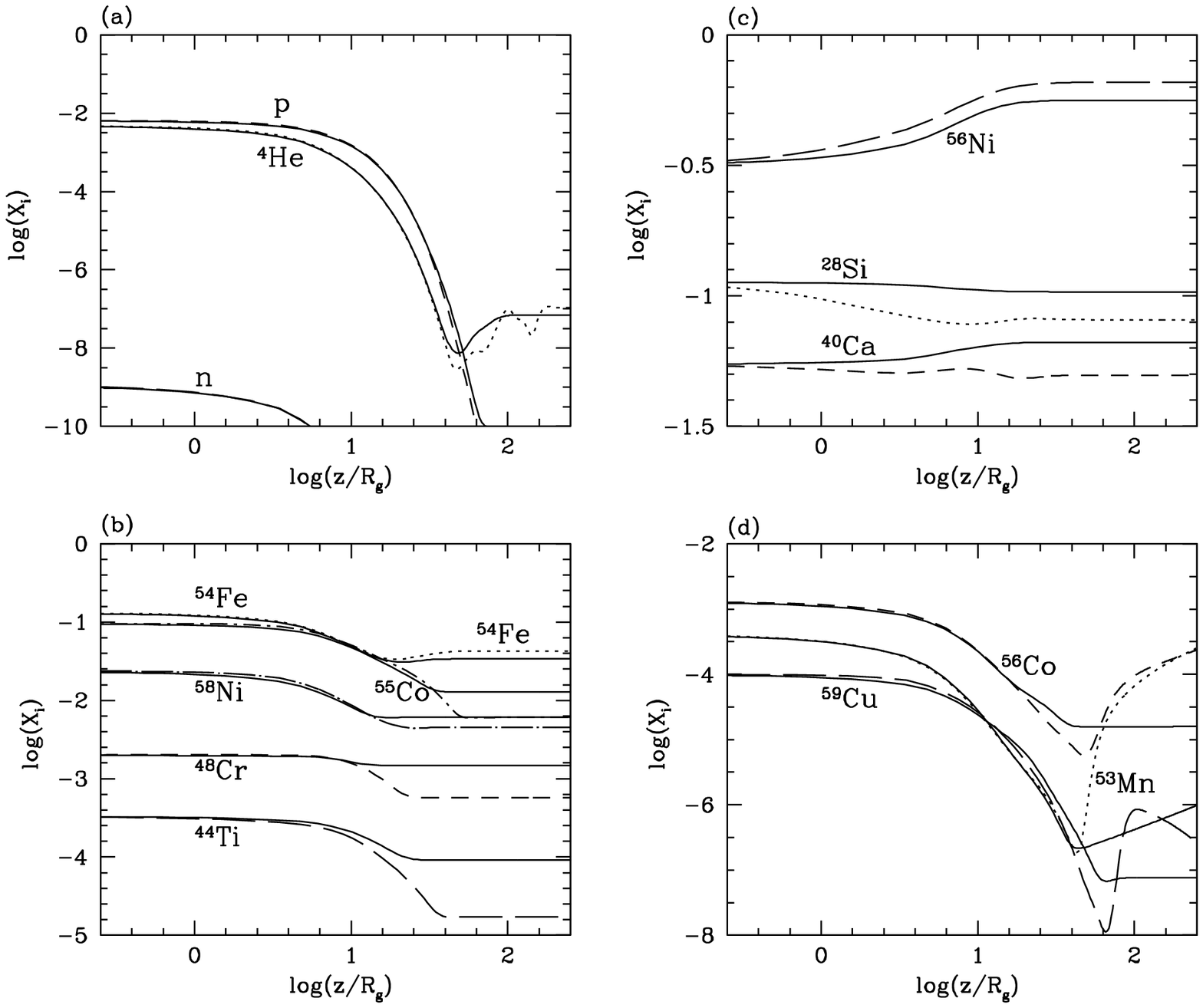}
\vskip-1.0cm
\caption{Same as in Fig. \ref{he40s}, except that $R_{ej}\sim 120R_g$.
}
\label{he120s}
\end{figure*}
This $R_{ej}$ lies in the zone overabundant in $^{56}$Ni,
$^{54}$Fe, $^{32}$S and $^{28}$Si.   
The entropy of the disk is calculated in the same way as done earlier and in this case it is $S \sim 24k$.
We assume that the outflow has, once the entropy same as that of the disk and then double
the entropy of the disk, and while considering each of these entropy cases we assume the same values for
$\dot{M}_{ej} / \dot{M}_{acc}$ as done earlier.
Figure 4. illustrates the evolutions of mass fraction of some of the most abundant elements in the outflow
when the outflow has the same entropy as that of the disk. 
%The convention of linestyles for the high 
%$v_{ej}$ and the low $v_{ej}$ cases are the same as before.
Although increasing the entropy decreases the density of the outflow, there is qualitatively not much 
change in the abundance evolution pattern for high entropy case 
compared to the low entropy case and hence we do not present a 
separate plot for the high entropy case in this situation.

We do not observe any trough-like feature in the mass fraction of $\alpha$-elements in this case.
Rather the elements which were abundant during the launching of the outflow almost maintain their high
abundance e.g., $^{28}$Si, $^{32}$S, $^{36}$Ar, $^{40}$Ca, $^{54}$Fe, $^{55}$Co, 
$^{56}$Ni, $^{57}$Ni and $^{58}$Ni etc.
The abundance evolution of some of these elements are plotted in Fig. 4. Whatever $^4$He was present in the 
disk to begin with, they 
recombine almost entirely to give rise to $^{56}$Ni. When $v_{ej}$ is low, the abundances of most of the 
elements
decrease more as the ejecta evolves away from the disk, compared to the high $v_{ej}$ case, because 
lower velocity of the ejecta ensures greater interaction among the elements which finally enables them to 
synthesize $^{56}$Ni. This explains why the mass fraction of $^{56}$Ni is higher when 
$v_{ej}$ is lower. 
This is one of the cases where many elements like $^{28}$Si, $^{32}$S, $^{36}$Ar, $^{40}$Ca, $^{54}$Fe, 
$^{55}$Co, $^{56}$Ni, $^{57}$Ni, $^{58}$Ni, $^{44}$Ti, $^{48}$Cr etc. survive in the outflow.
Thus, these elements should leave their signatures in the outflow. 
Here we emphasize that although 
emission lines of many of these heavy 
elements have been seen in the X-ray afterglows of several GRBs by XMM-Newton, Chandra, BeppoSAX etc., Swift is yet
to detect these lines. One possible resolution to this is that outflows may be mainly launched 
from the inner disk so that $^{56}$Ni is the primary component in it and the above mentioned elements
will thus not be present in outflows.

\subsection{Outflow from the Si-rich disk}

Now we analyze the abundance evolution in the outflow from the Si-rich disk, keeping other
parameters of the disk unchanged.

\begin{figure*}
%\vskip-5cm
   \centering
\includegraphics[scale=0.8]{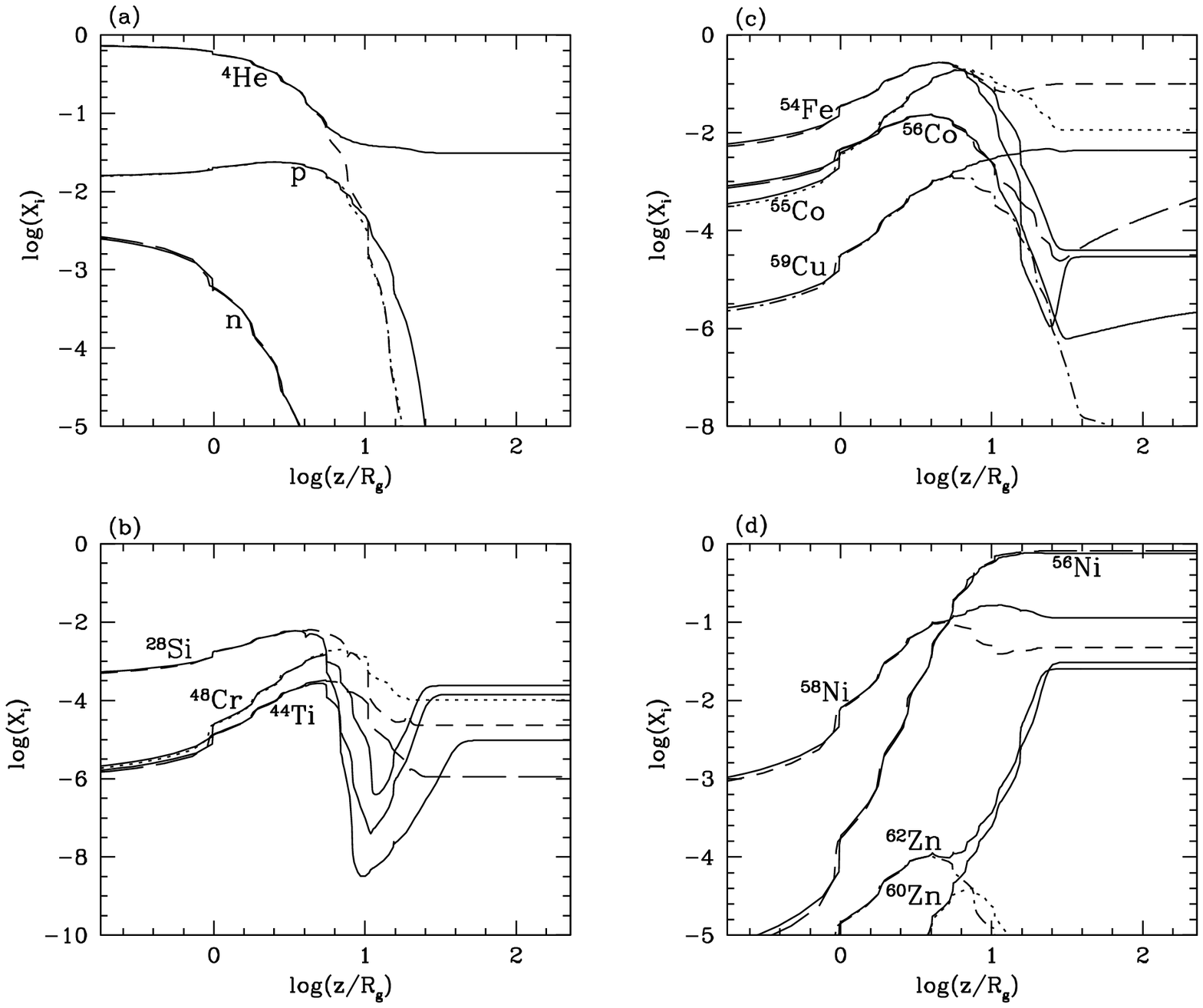}
\vskip-1.0cm
\caption{Abundance evolution of the ejecta while moving away from the disk, when outflow is being launched 
from $R_{ej} \sim 30 R_g $ of the accretion disk for a $3 M_\odot$ Schwarzschild black hole 
with $\dot{M}=0.001M_\odot s^{-1}$ with pre-SN Si-rich abundance at the outer disk and the entropy of the 
ejecta being same as that of the disk. The variation of mass fraction of the 
elements shown by the solid lines correspond to $\dot{M}_{ej} / \dot{M}_{acc} = 100$ and the same elements 
when shown by linestyles other than solid correspond to
$\dot{M}_{ej} / \dot{M}_{acc} = 0.1$.}
\label{si30s}
\end{figure*}
\newpage
\subsubsection{Outflow from $30R_g$}
%Here we consider outflow from $30R_g$ of the accretion disk surrounding the same $3 M_\odot$ Schwarzschild 
%black hole accreting at $\dot{M}=0.001M_\odot s^{-1}$ and having pre-SN Si-rich composition at the outer 
%disk.
$R_{ej} \sim 30R_g$ lies primarily in the He-rich zone of the disk. The entropy of the disk at this 
$R_{ej}$ is 
$S \sim 13k$. Here also we consider two cases, once we assume that the outflow has the same entropy as 
that of the disk and then double the entropy of the disk. For each of the cases we again assume that 
$\dot{M}_{ej} / \dot{M}_{acc} = 0.1$ and $100$. 

The abundance evolution of the elements, 
when the entropy of the outflow is high compared to that of the disk, have very similar trend to the case of
$R_{ej} \sim 40 R_g$ for He-rich disk. Hence we do not discuss the high entropy outflow from this radius 
of the disk in detail.  
However, the outflow having the same entropy as that of the disk has some different features.
When $\dot{M}_{ej} / \dot{M}_{acc} = 100$, i.e. $v_{ej}$ is high, as shown by Fig. 5,
the abundance evolutions of the elements resemble Fig. 1 qualitatively. Hence,
the underlying nuclear physics governing both the cases are similar. 
However, when $\dot{M}_{ej} / \dot{M}_{acc} = 0.1$, i.e. $v_{ej}$ is low, we find that the abundance evolution 
of the ejecta exhibits a markedly different trend. The most important difference is that, the 
$\alpha$-elements do not exhibit a trough in their abundance evolution pattern, 
although their mass fractions decrease slightly and then
become constant. Thus, in this situation the ($\alpha$,p) and ($\alpha$,n) reaction 
chains are not the dominant source for the synthesis of $^{56}$Ni, rather the direct recombination of 
$^4$He to $^{56}$Ni is more important, as a result of which $^4$He gets substantially depleted in this case,
which is quite clear from Fig. 1(a). The recombination of  $\alpha$-particles gains
so much of prominence because the $v_{ej}$ of the ejecta is the lowest in this case and 
the density of the ejecta is also high, as the entropy of the outflow is low (see, e.g., Fig. 3) which
favors greater recombination of $\alpha$-particles in the ejecta \cite{Fujimoto04, Witti}.

\subsubsection{Outflow from $75R_g$}
Here we study outflow from $R_{ej} \sim 75R_g$, when
the entropy chosen to be, once the same as that in the disk (which is $\sim 22k$) and then double 
than that in the disk. This $R_{ej}$ chiefly lies in the 
$^{54} \rm Fe/ ^{56}$Ni rich zone of the disk. 
However, $^4$He is also present abundantly in this part of the disk. The abundance evolutions of the 
elements in the ejecta resemble the 
abundance evolution trend of the elements in the outflow from $R_{ej} \sim 40R_g$ of the He-rich disk. 
Hence, we do not go into the detailed discussion of this particular case.  

\subsubsection{Outflow from $180R_g$}
\begin{figure*}
%\vskip-5cm
   \centering
\includegraphics[scale=0.8]{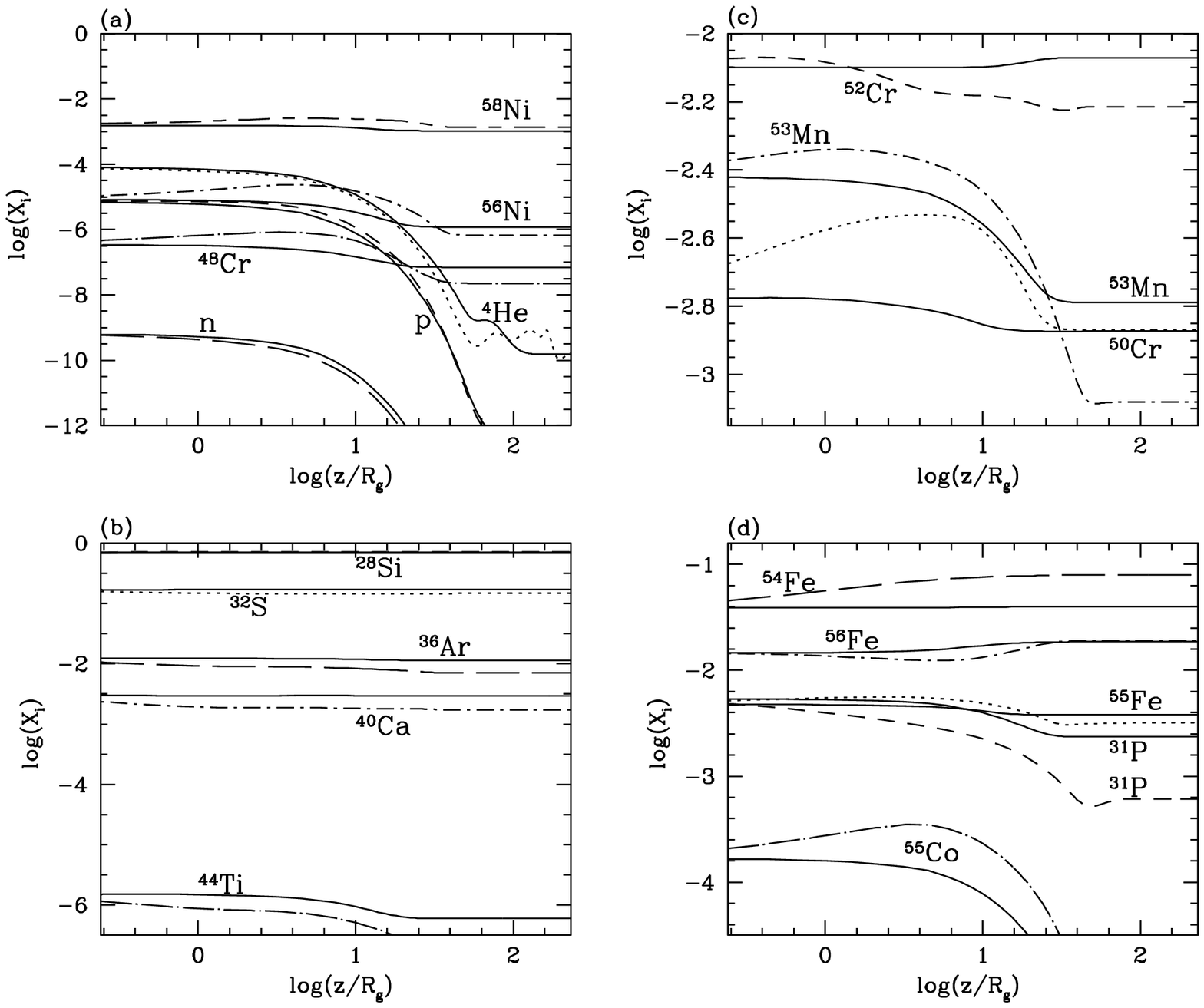}
\vskip-1.0cm
\caption{Same as in Fig. \ref{si30s}, except $R_{ej}\sim 180R_g$.
}
\label{si180s}
\end{figure*}

In this case we investigate outflow from $R_{ej} \sim 180R_g$.
When the entropy of the disk at $R_{ej}$ is $S \sim 28k$ and, as done before, we once assume that the outflow
has the same entropy as that of the disk and then assume that it has double the entropy of the disk.
This $R_{ej}$ lies chiefly in the $^{28}$Si/$^{32}$S rich zone of the disk. $^4$He is hardly present
in the disk during the launching of the outflow.
The abundance evolution of various elements present in the outflow is shown in Fig. 6. 
The outflow from this particular $R_{ej}$ almost retains the composition of the disk. 
We note that 
this is one case where $^{56}$Ni is hardly synthesized in the outflow and hence outflow from this $R_{ej}$ 
will not lead to an observable supernova.
Elements chiefly present in the outflow are $^{28}$Si, $^{32}$S, $^{36}$Ar, $^{40}$Ca, $^{54}$Fe, 
$^{56}$Fe and $^{52}$Cr and their mass fractions remain roughly constant throughout the outflow.
%Emission 
%lines of most of these elements have been observed in the X-ray afterglows of several GRBs.

\subsection{Outflow from the disks with high accretion rate}

\begin{figure*}
\vskip 0.8in
   \centering
\includegraphics[scale=0.85]{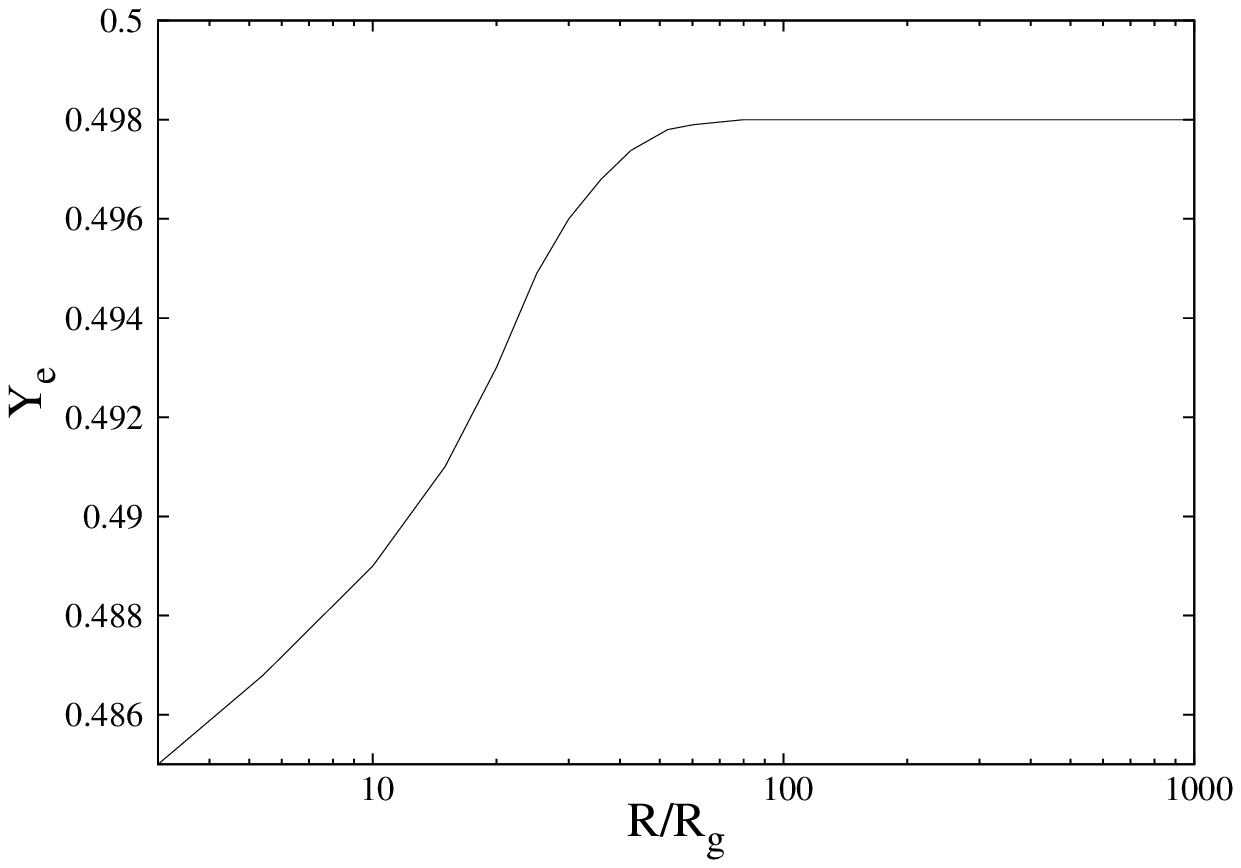}
%\vskip-0.2cm
\caption{Evolution of the electron fraction inside the disk with $\dot{M}=0.01M_{\odot} s^{-1}$ 
irrespective of the initial abundance.}
\label{tempq1p9}
\end{figure*}

\begin{figure*}
\vskip -1.8in
   \centering
\includegraphics[scale=0.85]{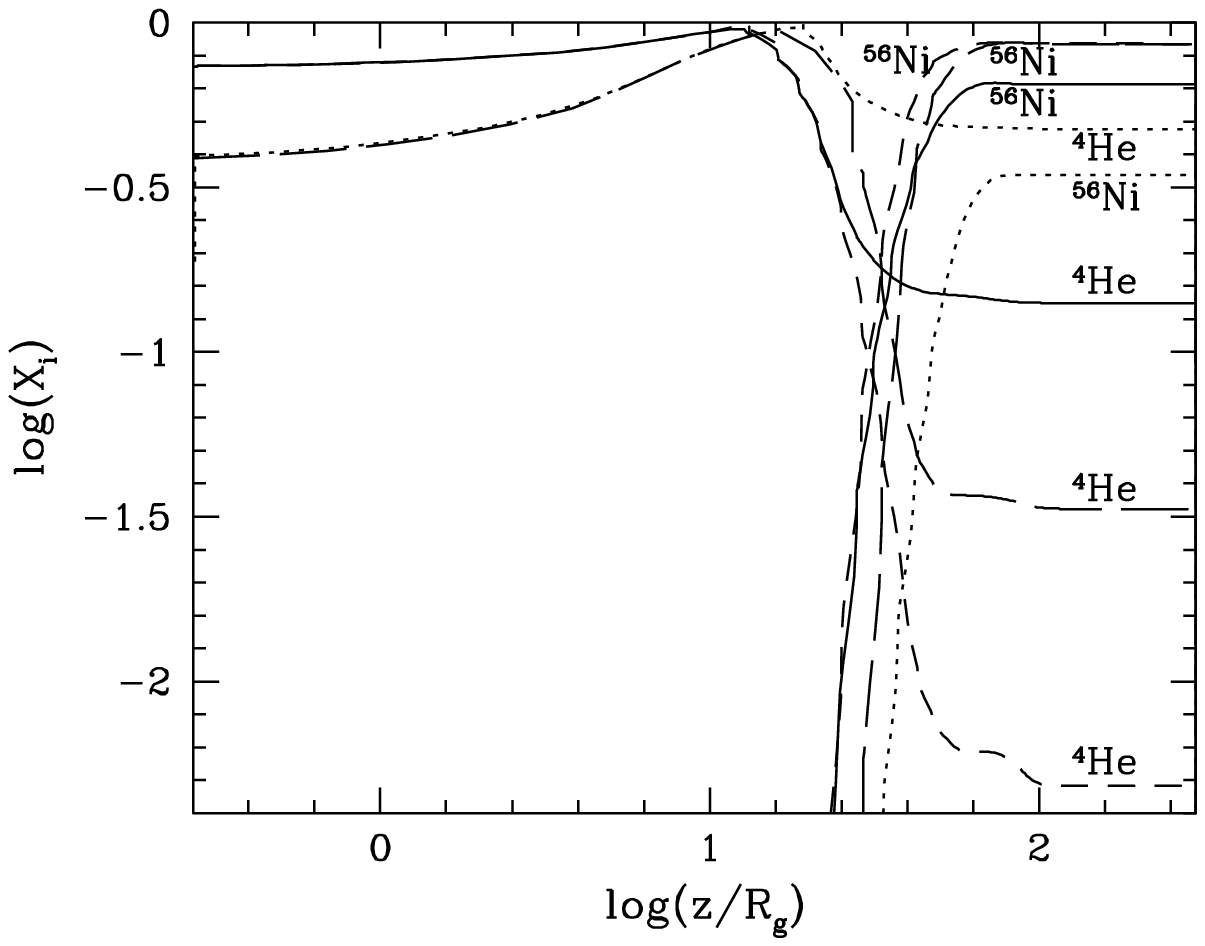}
\vskip-3.2cm
\caption{Yields of $^{56}$Ni and $^{4}$He in the outflow which is being launched from
$R_{ej} \sim 80 R_g$ of the accretion disk for a $3 M_\odot$ Schwarzschild black hole 
with $\dot{M}=0.01M_\odot s^{-1}$ with pre-SN Si-rich abundance at the outer disk. 
The solid lines correspond to high velocity and low entropy cases, dotted lines
correspond to high velocity and high entropy, short-dashed lines correspond to low velocity and low entropy 
and long-dashed lines correspond to low velocity and high entropy cases. }

\label{tempq1p9}
\end{figure*}

Here we consider outflow from the accretion disks surrounding a $3 M_\odot$ Schwarzschild 
black hole accreting at $\dot{M}=0.01M_\odot s^{-1}$. In these disks, inside $500 R_g$, the 
abundance evolutions of the elements start appearing identical irrespective of the initial 
abundance at the outer disk. The region $90R_g \lesssim R \lesssim 500R_g$ forms the He-rich zone of the 
disk. 
For $R \lesssim 90R_g$ the $\alpha$-particles break down into neutrons and protons. 
Thus, in the innermost region of the disk, as is evident from Fig. 7, the electron fraction $Y_e$ gets
reduced from the value $0.5$ (unlike the case of the disk with $\dot{M}=0.001M_\odot s^{-1}$). 
Moreover, in the outflow from this region, the neutrons and protons recombine
to form alpha particles and further away from the disk the alpha particles
recombine among themselves to ultimately form $^{56}$Ni \cite{Surman11}.
Hence, if we consider outflows from $R_{ej} < 200 R_g$
of both the He-rich and Si-rich disks, the abundance evolutions of the elements
in the outflow will exhibit similar behavior.

As an example, we consider outflow from $R_{ej} \sim 80R_g$ of the Si-rich disk. 
Thus, to begin with, the outflow had plenty of $\alpha$-particles apart from substantial amounts of 
free nucleons. As 
before $^{56}$Ni is abundantly synthesized at the expense of $^{4}$He. The yields of $^{56}$Ni and 
$^{4}$He in the outflow by varying the entropy and velocity of ejection of the outflow are
illustrated in Fig. 8. Of all the four cases considered here, the mass fraction of $^{56}$Ni is highest and 
$^{4}$He is lowest when the outflow has low entropy and low velocity of ejection. This is because when the
entropy is low, the density is high (see, e.g., Fig. 3) and the low velocity of expansion in the dense 
outflow medium favors
greater recombination of $\alpha$-s to $^{56}$Ni. 
Note that lower $v_{ej}$ is a more important factor than higher density for the conversion of $\alpha$-s to 
$^{56}$Ni and hence next to the case with low entropy and low $v_{ej}$, the yield of $^{56}$Ni is higher 
when $v_{ej}$ is low but entropy is high. 
%For this case the abundance of $^{56}$Ni and $^{4}$He is shown 
%by long-dashed lines. The mass fraction of $^{56}$Ni and $^{4}$He when $v_{ej}$ is high and entropy is low 
%is shown by solid lines and when $v_{ej}$ is high and entropy is
%high is shown by dotted lines. 
The relative yields of $\alpha$-s and $^{56}$Ni in the remaining two cases 
can be explained in similar fashions.

\newpage
\section{Discussions and Conclusions} 
We have studied nucleosynthesis in outflows from accretion disks associated with the fallback 
collapsars using already established outflow model and the nuclear reaction network. 
While studying the abundance evolution in the outflow, we have considered different velocities of expansion 
and  
entropies of the outflow. We once considered that the entropy of the outflow is same as that of the disk
which is the case for an MHD-driven outflow and subsequently assumed that the entropy is double than that 
of the disk,
because the entropy of the outflow may get enhanced due to viscous heating.
The entropy of the outflow is mostly raised when it is being launched from the disk and after that it
remains roughly constant \cite {Fujimoto04, Pruet04}.
We have found that the nucleosynthesis products do not change by varying the entropy of 
the outflow. In fact, the abundance evolution patterns of the elements also do not change qualitatively
by changing the entropy of the outflow, although the final abundance of the elements may slightly vary.

We have shown in our previous works (Banerjee \& Mukhopadhyay 2013a,b) that the accretion disks formed by 
the Type II 
collapsars have several zones characterized by dominant elements and we have considered outflows from each 
of these zones. Outflow from the He-rich and the $^{54}$Fe-$^{56}$Ni-rich zones of the disk always leads to 
the synthesis of copious amounts of $^{56}$Ni. The synthesis of $^{56}$Ni is important because it signifies 
that the outflow will drive a supernova explosion.
Apart from $^{56}$Ni, the isotopes like $^{57}$Ni, $^{58}$Ni, $^{59}$Cu, $^{44}$Ti, $^{60}$Zn 
and $^{62}$Zn are also 
synthesized 
abundantly in the outflow from these zones of the disk (see, for e.g., Fig. 1). 
If the velocity of ejection is higher than a certain threshold, the $\alpha$-elements in the outflow from 
the aforementioned zones of the disk exhibit a trough like feature in their abundance evolution pattern. 
However, if the velocity of expansion is too low then this feature vanishes. The reason is explained in 
great detail in \S 4.1.1 \& 4.2.1 and we are reporting this very feature for the first time in the 
literature. Although we are not yet aware of any observational consequences of this feature, it
certainly bears significance in terms of the underlying nuclear physics governing it.
It reveals that certain chain of nuclear reactions becomes 
important under the specific hydrodynamic conditions of the outflow and therefore is important in the 
context of nuclear astrophysics.
We have shown that if we consider outflow from the Si-rich zone of the disks, the abundance in the outflow
is not much changed from that of the disk, i.e, the outflow remains rich in silicon. 
There may be stellar explosions in these cases, but since there is no $^{56}$Ni to begin 
with, it cannot decay to $^{56}$Fe and hence there is nothing to power the supernova light curve. Thus 
there will be no observable supernova.

We expect that those elements which survive in the outflow will leave their signatures in the observations.
%These ejected elements presumably collide with the material in the interstellar medium, lose their kinetic 
%energy, and emit less energetic radiations like X-rays, visible light and radio waves which form the 
%afterglow. These afterglows can persist for months as the energies gradually shift to lower frequencies. 
Although
emission lines of many of these elements have been discovered in the X-ray afterglows of GRBs
by XMM-Newton, BeppoSAX, Chandra etc., e.g., iron lines 
in GRB 970508 (Piro et al. 1999), GRB 970828 (Yoshida et al. 1999) and GRB 000214 (Antonelli et al. 2000); 
magnesium, silicon,
sulphur, argon, calcium lines in GRB 011211 (Reeves et al. 2003), these results should be taken with 
caution
as a more recent satellite Swift has not detected these lines yet (Zhang et al. 2006, Hurkett et al. 2008).
Nevertheless, if we observe these elements in the outflow, we can have an idea about the nature of the
accretion disks from where they are ejected, based on the present computations. 

It has been recently speculated that the Ultra High Energy Cosmic Rays (UHECRs) may be composed of heavy 
nuclei 
like $^{56}$Ni. The Yakutsk data analysed the muon component of the UHE air showers and reported the 
presence of heavy nuclei component in the UHECR spectrum \cite{Glushkov07}. Moreover, the elongation rate 
measurements done by the Pierre Auger Observatory team indicate the possible presence of heavy or 
intermediate mass nuclei in the UHECRs. Wang et al. (2008) explored under what conditions these UHE heavy 
nuclei
synthesized in a GRB scenario survive photodisintegration and become a major component of UHECRs.
They investigated the survival of these heavy nuclei both in the context of internal shock and external
shock GRB scenarios. In the internal shock scenario, the outflows, which are collimated as jets, 
themselves are the sources of heavy nuclei, whereas in the situation of external shocks the jet interacts 
and entrains heavy nuclei as it passes through the surrounding interstellar medium. 
In the present work, we show to have synthesized copious amounts of $^{56}$Ni and other heavy elements in 
the outflow. These heavy elements once synthesized in the outflow spread out and contaminate the 
interstellar medium. Therefore, if the supernovae we discussed are associated with GRBs,
there are possibilities that the nuclei synthesized in the outflows resulting in stellar explosions may
be entrained with the jet of material in the burst and be plausible components of UHECRs.
Note that Metzger et al. (2011) discussed the possibility of heavy nuclei synthesized in the 
outflows from GRBs (including disk winds) as components of UHECRs.
This possibility was further studied by Horiuchi et al. (2012).

Metzger (2012) studied one-dimensional models of nuclear burning in accretion disks with outer composition similar to 
those of collapsars and $\dot{M}$ similar to those considered in this work $\sim 0,001 M_{\odot} s^{-1}$.
He found that the outer region of the disk could be thermally unstable, due to high nuclear energy
generation. Hence, if the disk is unstable, then the steady flow assumption for either disk or outflow 
is not valid. This work was followed up by two-dimensional numerical simulations by Fernandez \& Metzger (2013), 
where the authors actually found that detonations could occur in the disk miplane, possibly unbinding 
the disk material dynamically.
However, these concerns are not so important in our case, when we find that the viscous energy is at least
two orders of magnitude higher than the nuclear energy even in the outer regions.
This may be due to the 
difference in the disk model for our case (Banerjee \& Mukhopadhyay 2013a) compared to the case of Metzger 
(2012). Moreover, Metzger (2012) considered disks formed by mergers of white dwarfs
with neutron stars or black holes, while we discuss about collapsar
accretion disks. Hence, the temperature and the density profiles of the two disk models
are expected to be different, which might affect the nuclear energy generation.
Further, the central object in our case is always a stellar mass black hole and not a neutron star, 
which brings in a deeper potential well and potentially lower angular momentum (and hence 
smaller outer radius) in our disks. Finally, in the inner region of the disk, 
the nuclear reactions become endothermic. All of them argue for stable collapsar disks. 
Nevertheless, one should check with the possible
nuclear instability (as was done by one of us in the context of low density disks,
Mukhopadhyay \& Chakrabarti 2000, 2001) in collapsar disks itself based on different disk models, which might 
prevent us from choosing steady accretion flows. 

We now roughly estimate the change in the mass fraction
of a particular species  due to these supernova events during the lifetime of a galaxy.
If we assume that the mass of the galaxy where these supernova events take place is typically about 
$10^{11} M_{\odot}$, the typical age of the galaxy is approximately $10^{10}$ years, during the lifetime of 
the galaxy on an average one supernova event takes place in every $100$ years and the
typical change in the mass fraction of the i-th species during one such supernova event is $10^{-3}$,
then the average change of mass fraction of the i-th species in entire galaxy in its lifetime can be given
by  
\begin{equation}
\langle \Delta X_i \rangle = 10^{-6}\left(\frac{\Delta X_i}{10^{-3}}\right)\left(\frac{100}
{T_{S}}\right)\left(\frac{T_{G}}{10^{10}}\right)\left(\frac{10^{11}}{M_G}\right).
\end{equation}
%If we now consider the total change in the mass fraction $\Delta Y_i$ of the i-th species in the entire 
%universe which comprises of, say, $N$ galaxies of equal mass, then we can show that 
%$\Delta Y_i =\Delta X_i $. 

As an example let us consider $^{60}$Zn. Of all the cases considered here
the mass fraction of $^{60}$Zn at largest distance from the accretion disk is maximum when the 
velocity of ejection is high and the entropy of the ejecta is same as that of the He-rich accretion disk 
at $R_{ej} \sim 40 R_g$. The 
maximum mass fraction of 
$^{60}$Zn in the outflow is $\sim 0.0365$. If we put $\Delta X_i=0.0365$ in 
equation (8) with $T_S=100$, $T_G=10^{10}$ and $M_G=10^{11}$, we obtain $\langle \Delta X_i \rangle 
=3.65 \times 10^{-5}$. This gives the maximum contamination of $^{60}$Zn in the galaxy.
Similarly for $^{28}$Si, $^{32}$S, $^{59}$Cu, $^{48}$Cr and  $^{44}$Ti the value of $\langle \Delta 
X_i \rangle$ obtained are $7 \times 10^{-4}$, $1.7 \times 10^{-4}$, $6 \times 10^{-6}$, $1.5 \times 
10^{-6}$ and $7 \times 10^{-7}$ respectively.
If the mass of 
the galaxy is chosen to be smaller, say $10^9M_\odot$, then above values of
$\langle \Delta X_i \rangle$ will increase by two orders of magnitude which appear quite significant.

We can proceed in the similar fashion for other elements as well. For every element we have to 
consider the outflow model giving rise to the maximum mass fraction of the element generally away from the 
disk. This gives an upperbound of the contamination of the element in the galaxy during its lifetime. 

We now attempt to estimate the possible mass of various elements ejected in the outflow which will 
eventually 
enable us to predict whether the emission lines of these elements should be observed or not in the 
afterglow.
In a collapsar II accretion disk, $\dot{M}$ of $10^{-4}-10^{-2} M_{\odot}s^{-1}$
is approximately maintained for hundreds to thousands of seconds
(MacFadyen et al. 2001). 
In our case, when matter is being accreted in the disk at the rate of $0.001 
M_{\odot}s^{-1}$, the accretion continues for approximately $100 s$.
Thus the total amount of matter supply is $0.1 M_{\odot}$. A part of this matter is ejected from
the disk producing outflow. Hence, we can calculate the mass of the various
individual elements like  $^{44}$Ti, $^{48}$Cr, $^{59}$Cu, $^{60}$Zn, $^{62}$Zn etc.
present in the outflow. We estimate, as an example, the mass of $^{62}$Zn in the outflow.

$^{62}$Zn is synthesized abundantly in the outflow from $40 R_g$ of the He-rich
disk (see Fig. 1 and Fig. 2). We evaluate the amount of $^{62}$Zn being 
ejected in the outflow from Fig. 1. 
In the outflow, we have already chosen that $v_{ej}$ is once $\sim 10^8 \rm cm ~ s^{-1}$ and then $\sim 
10^5 \rm cm~s^{-1}$. 
When $v_{ej}$ is high, $\dot{M}_{ej}/\dot{M}_{acc} \sim 100$.
If $\dot{M} = \dot{M}_{ej} + \dot{M}_{acc} =0.001 M_{\odot}s^{-1}$, then $\dot{M}_{ej} \sim 
9.9 \times 10^{-4} M_{\odot}s^{-1}$. If we assume that the outflow continues for $100 s$, 
i.e, as long as steady accretion is maintained, then we obtain an upper limit of the total amount of mass 
ejected $\sim 0.099 M_{\odot}$. Now the maximum mass fraction of $^{62}$Zn 
as the ejecta evolves away from the disk is $\sim 0.0082$ (see Fig. 1). 
Thus the amount of $^{62}$Zn ejected would be $8 \times 10^{-4} M_{\odot}$.

When $v_{ej}$ is low, $\dot{M}_{ej}/\dot{M}_{acc} \sim 0.1$. 
In that case, for $\dot{M} = 0.001 M_{\odot} s^{-1}$, $\dot{M}_{ej} \sim 9.09 \times 10^{-5} 
M_{\odot}s^{-1}$.
Assuming, as before, that the outflow continues as long as accretion takes place, i.e,
for $\sim 100 s$, the upper limit of the total amount of matter ejected 
in this case is $\sim 0.009 M_{\odot}$. The maximum mass fraction of $^{62}$Zn as the ejecta evolves away
from the disk is $\sim 0.0035$ (as is evident from Fig. 1). Thus, the the total mass of 
$^{62}$Zn ejected in this case is $3.15 \times 10^{-5} M_{\odot}$.

From similar calculations, we obtain the amount of $^{44}$Ti,  $^{48}$Cr, $^{59}$Cu,
$^{60}$Zn, $^{32}$S and $^{28}$Si in the outflow to be $7 \times 10^{-5}$, $1.5 \times 10^{-4}$, $6 \times 
10^{-4}$, $4 \times 10^{-3}$, $1.7 \times 10^{-2}$ and $6.9 \times 10^{-2}  M_{\odot}$
respectively when $v_{ej}$ is high, some of which may be sufficient
to be observed in the emission lines. However, 
for low $v_{ej}$, apart from $^{28}$Si, $^{32}$S and $^{60}$Zn the masses of all the other elements are 
much less and hence not sufficient to be
observed in the emission lines, as will be argued below. 
The masses of $^{28}$Si, $^{32}$S and $^{60}$Zn in the outflow when $v_{ej}$ is low are $6.5 \times 
10^{-3}$, 
$1.35 \times 10^{-3}$ and $1 \times 10^{-4} M_{\odot}$ respectively.

According to Fujimoto et al. (2004), the amount of matter ejected from 
the disk in a Type II collapsar can go upto $0.1 M_{\odot}$, depending 
on the amount of explosion energy in the mild supernova explosion associated with it. 
According to them, if the explosion energy is $\gtrsim 1 \times 10^{51} ergs$,
the ejecta mass will be less than $0.01 M_\odot$ and if the explosion energy is less 
than $1 \times 10^{51} ergs$, the yields in the outflow will be $\sim 0.1 M_\odot$. Although we 
have not calculated the explosion energy, we have obtained similar masses of 
the ejecta from our calculations above (e.g. $0.009 M_\odot$ and $0.099 M_\odot$ when $v_{ej}$
in low and high respectively). 

Lazzati et al. (1999) evaluated the iron mass required to obtain an emission line of iron
in the GRB afterglow. According to them, the required amount of iron mass $M_{Fe} \gtrsim 2.3 \times 
10^{-5}F_{(Fe,-13)} t_5 R^{2}_{16}/ (q E_{52}) ~ M_{\odot}$. 
See Lazzati et al. (1999) for the definition of various symbols.
Assuming $F_{(Fe,-13)} \sim 1$ (Lazzati et al. 1999), 
$q \sim 0.1$ at most (Ghisellini et al. 1999), and the remaining parameters in the expression
of $M_{Fe}$ to be $\sim 1$ (Lazzati et al. 1999), the required mass of iron should be 
$\sim 2.3 \times 10^{-4} 
M_{\odot}$.
If this is roughly true for other elements, 
%then we can generalize the expression for $M_{Fe}$ 
%for any element replacing 
%the line flux $F_{(Fe,-13)}$ by $F_{(Z,-13)}$, where $Z$ represents  
%the desired element. If $F_{(Z,-13)} \sim 1$ and the values of the other parameters remain the same as before,
then our model predicts that there is a possibility of observing silicon, sulphur, zinc, copper and 
chromium lines  
in the GRB afterglow as the masses of these elements, as calculated above, represent a value
$\gtrsim 2.3 \times 10^{-4} M_{\odot}$.

\section*{Acknowledgments}

\indent \indent This work was partly supported by the ISRO grant ISRO/RES/2/367/10-11.

\end{document}